# The hot pick-up technique for batch assembly of van der Waals heterostructures

Filippo Pizzocchero,[1] Lene Gammelgaard,[1] Bjarke Sørensen Jessen,[1,2] José M. Caridad,[1] Lei Wang,[3] James Hone,[4] Peter Bøggild,[1,*] Timothy J. Booth[1]

[1] DTU Nanotech, Technical University of Denmark, Ørsteds Plads 345C, Kongens Lyngby, Denmark

[2] Center for Nanostructured Graphene (CNG), Technical University of Denmark, Ørsteds Plads 345C, Kongens Lyngby, Denmark

[3] Kavli Institute (KIC), Cornell University, 14853 Ithaca, USA

[4] Department of Mechanical Engineering, Columbia University, 10027 New York, USA

* Contacts: peter.boggild@nanotech.dtu.dk

**Abstract**

The assembly of individual two-dimensional materials into van der Waals heterostructures enables the construction of layered three-dimensional materials with desirable electronic and optical properties. A core problem in the fabrication of these structures is the formation of clean interfaces between the individual two-dimensional materials which would affect device performance. We present here a technique for the rapid batch fabrication of van der Waals heterostructures, demonstrated by the controlled production of 22 mono-, bi- and trilayer graphene stacks encapsulated in hexagonal boron nitride with close to 100% yield. For the monolayer devices we found semiclassical mean free paths up to 0.9 µm, with the narrowest samples showing clear indications of the transport being affected by boundary scattering. The presented method readily lends itself to fabrication of van der Waals heterostructures in both ambient and controlled atmospheres, while the ability to assemble pre-patterned layers paves the way for complex three-dimensional architectures.

**Introduction**

The controlled isolation and assembly of single- and few-layer sheets of two-dimensional (2D) materials into van der Waals (vdW) heterostructures has thrown open the doors for the design and fabrication of new devices and functionalities based on 2D materials with an unprecedented flexibility and atomic precision. The perspectives are astounding, with applications ranging from electronics,[1, 2] photovoltaics,[3, 4] and sensing[5, 6] through to memory storage[7] with the intriguing possibility of increased performance and a wide range of functionalities on flexible, transparent substrates.[8, 9]

Clean interfaces between 2D materials result in the best device performance [10-15] – any contaminants present between interfaces usually gather up as 'blisters' between layers, leading to deterioration of transport properties[16] as well as compromising the perfect van der Waals interlayer adhesion between the layers. Devices can be produced in controlled atmospheres such as in gloveboxes[17] but without special cleaning steps to remove ad-layers and careful monitoring, contamination and decreased interlayer adhesion may still be an issue. Oxygen plasma pre-treatment increases the number and area of individual flakes of 2D materials produced by micromechanical exfoliation on oxidised silicon substrates,[18] but results in increased substrate interaction, which dopes graphene and is manifested as a reduced Raman $I(2D)/I(G)$ peak ratio.[19]

We present here room temperature mobility measurements from a batch-fabricated set of 22 mono-, bi and trilayer blister-free encapsulated graphene field effect devices - assembled in ambient atmosphere - with over 280 individual metal-graphene contacts. We find that blisters of trapped interfacial contamination commonly observed in such samples by optical and atomic force microscopy can be completely eliminated by stacking individual 2D crystals into van der Waals heterostructures at temperatures of 110°C, even in ambient atmosphere. We prove that the reduction of the I(2D)/I(G) ratio in graphene due to plasma treatment of the cleavage substrate is fully reversed when the crystals are lifted from this substrate, meaning that plasma pre-treatment is a viable way to obtain large and pristine 2D material flakes for integration into van der Waals heterostructures. By actively tuning the interfacial adhesion and cleanliness through temperature whilst completely avoiding any contact with liquids in the stacking procedure, we are able to controllably pick up and drop down 2D materials, including single layer crystals that have been pre-patterned using electron beam lithography (EBL). This method enables us to produce a statistically significant dataset of field effect mobility measurements from 22 mono-, bi- and trilayer encapsulated graphene devices with more than 280 contacts. 7 of the 16 monolayer devices and 55% of the measurements display carrier mean free paths comparable or exceeding the channel width, with carrier mean free paths limited by boundary scattering.[20] Bi- and trilayer devices show diffusive behaviour with average mobilities above 20,000 cm$^2$/Vs and 15,000 cm$^2$/Vs, respectively. No annealing at high temperatures is necessary to obtain this high performance.[13, 21, 22]

**Results**

*Exfoliation of 2D materials and Raman precharacterisation*

Mono-, bi- and trilayer graphene and thin (thickness in the range 20±10 nm) hexagonal boron nitride (hBN) flakes, all with areas of over 2000 µm$^2$, are produced on oxidised silicon substrates by micromechanical cleavage. A combination of oxygen plasma pre-treatment of the substrate and thermal release of the adhesive tape[18] is employed to increase the size and number of the crystals produced - exfoliation on non-plasma treated substrates yields smaller and fewer crystals (Figure 1 a). Plasma treatment is known to introduce trapped charges in oxidised silicon, while heating increases substrate conformity and induces roughness in graphene[23]. Both of these effects can potentially impact device performance. The Raman *I(2D)/I(G)* peak ratios of graphene flakes cleaved on plasma treated oxide show a noticeably higher doping than for graphene on untreated oxide (*I(2D)/I(G)* peak ratio = 1.1 on plasma treated vs. 1.8 on untreated - Figure 1 b). This trend is not present in the same graphene flakes when lifted from the substrate using hBN crystals. For such samples, the *I(2D)/I(G)* peak ratio increases from less than 2 to 4.2 and 3.3 for graphene from plasma treated and untreated SiO$_2$ respectively - Figure 1 b. This shows that plasma cleaning of the substrate can be used to produce more and larger 2D crystals, and has no permanent impact on those crystals after lifting them from the production substrate.

*Heterostructure assembly*

We use a polypropylene carbonate (PPC) coated polydimethylsiloxane (PDMS) block mounted on a glass slide to capture (pick-up) and release (drop-down) 2D materials (Figure 1 d, e). For full details of the described techniques, see Supplementary Methods. The slide and polymer block is positioned in *x*, *y* and *z* using a micromanipulator (Figure 1 c) fixed to a heated microscope stage. By tuning the temperature above the boiling point of water and the glass transition temperature of the polymer, flakes can be reproducibly picked up or dropped down at desired positions (Figure 1 f, Figure 2 a-c). Higher temperatures during transfer are enabled by the use of oxygen plasma treatment of the PDMS prior to application of the PPC layer, which increases the adhesion between the PPC and the PDMS and prevents delamination at high temperatures.

Temperatures above 110°C favour the van der Waals adhesion between hBN and graphene, whereas a temperature of 40°C is sufficient to lift hBN from oxidised silicon cleavage substrates using the PPC/PDMS block. We note that assembly at high temperatures produces adhesion forces between graphene and hBN which are strong enough to tear the graphene flake, leaving behind graphene regions which are not covered by hBN adhering to the oxidised silicon substrate (Figure 2 a-d). Such high adhesion forces lead to a pick-up yield of close to 100%. It is always possible to pick up hBN from $SiO_2$ using PPC, regardless of the temperature, but we are unable to pick up graphene using PPC in the same way – graphene adheres more strongly to the oxidised silicon surface than the PPC, leading to the tearing of edges that can be seen in Figure 2 d. Such structures are a result of the reported preferred tearing directions in monocrystalline graphene.[24] A temperature of 110°C is required to pick up graphene from plasma-treated oxidised silicon using an hBN crystal on PPC – at 40°C the graphene adhesion to the plasma-treated $SiO_2$ surface dominates. For graphene on $SiO_2$ surfaces that have not been plasma treated, hBN can be used to pick up the graphene flakes regardless of temperature, as previously reported.[10] Lithographically pre-patterned 2D materials can also be reliably picked up for subsequent encapsulation in this way (Figure 1 e - h), which allows a greater flexibility in device architecture.

A key feature of this technique is the ability to produce large numbers of encapsulated graphene samples in parallel, by covering many graphene flakes on their production substrate with dropped down hBN flakes (Figure 1 f i-iii), subsequently picking these hBN/graphene stacks up, and finally dropping them down on new hBN flakes produced on oxidised silicon substrates (Figure 1 f iv-vii). This enables the rapid batch fabrication of many heterostructure stacks such as those shown here, and the parallelization of subsequent processing steps of lithography and etching to define devices.

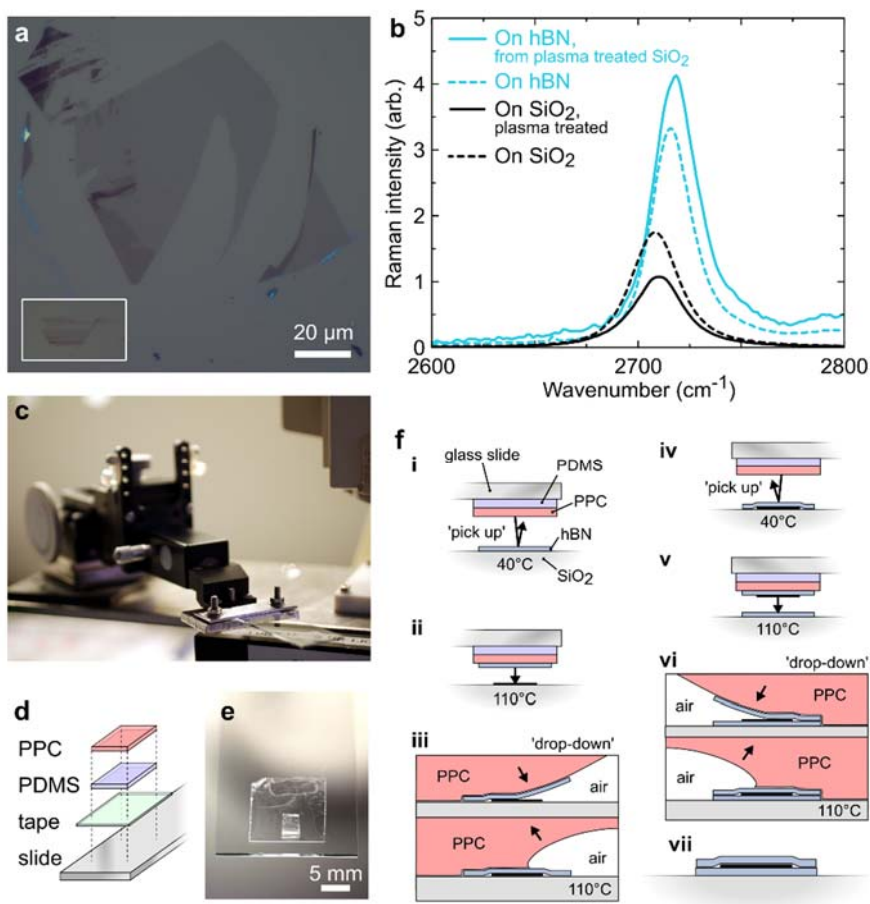

*Figure 1: Assembly and precharacterisation of heterostructures. a Monolayer graphene flakes produced on oxygen plasma treated $SiO_2$. Inset (white square): typical graphene flake size on non-plasma treated $SiO_2$. To scale with (a) b Graphene Raman 2D peak dependence on substrate – the Raman I(2D)/I(G) ratio increases for graphene picked up with hBN from $SiO_2$ irrespective of the oxygen plasma pre-treatment of the substrate. c Micromanipulator with slide assembly used for assembly of heterostructures. d, e polymer*

*stack on glass slide used for pick-up and drop-down. **f** schematic process flow for assembly of 2D heterostructures by pick-up and drop-down.*

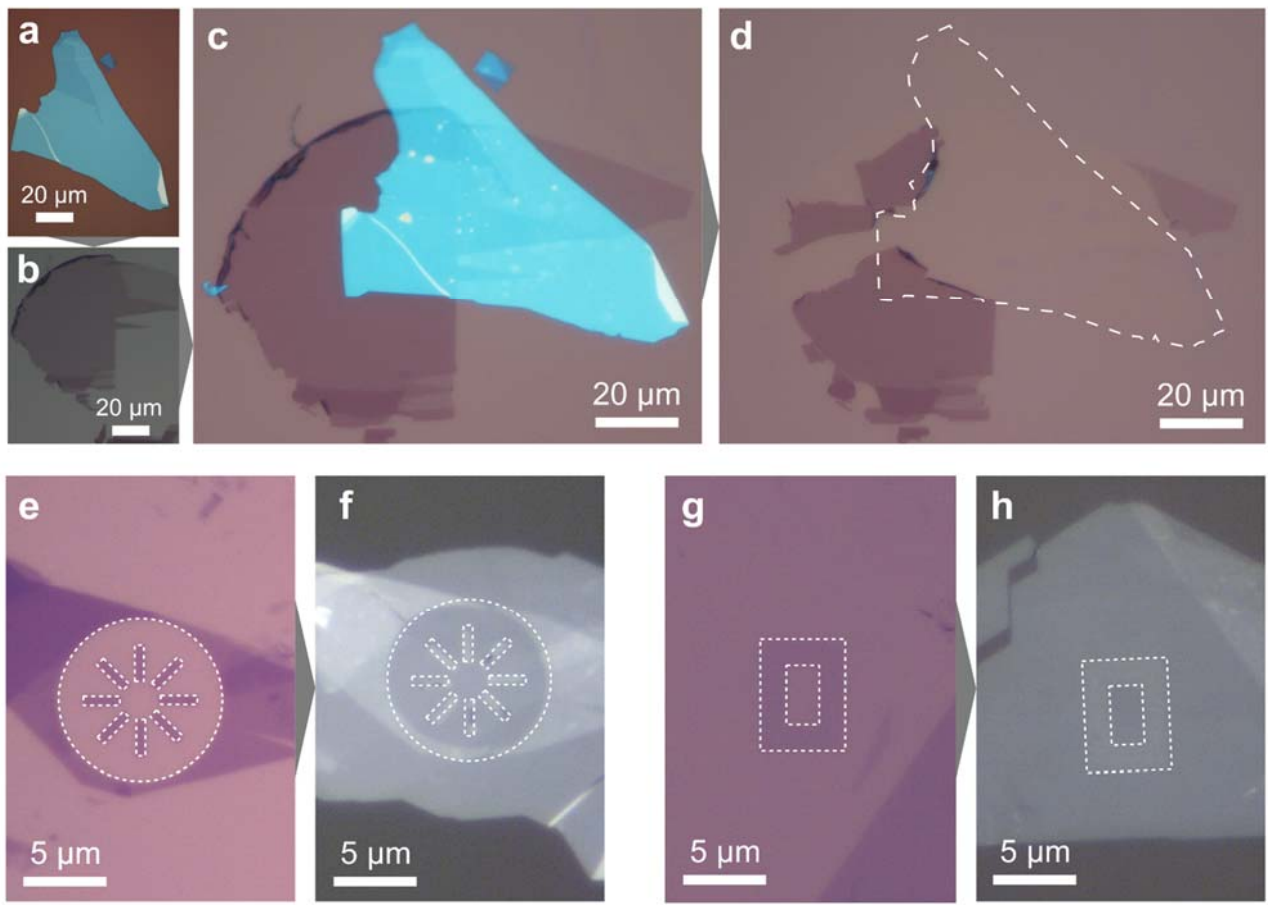

*Figure 2: Adhesion between pristine and patterned 2D materials. **a** hBN flakes are dropped down onto **b** graphene flakes (single and bilayer shown) to produce **c** a stack. **d** Interlayer adhesion between the hBN and graphene is sufficient to selectively tear the graphene away from the substrate. The dashed line indicates the previous extent of the hBN flake before pick-up. **e-h** Graphene layers which have been pre-patterned with electron beam lithography (e, g) can be picked up using hBN to produce laterally patterned vdW heterostructures (j, h).*

### Characterising and avoiding blisters during assembly

We compare the number and size of blisters formed during stacking at 110°C versus 40°C using optical and atomic force microscopy, and find that the cleanliness of the interface critically depends upon both the speed of the approach of the flakes and the temperature. One point of the PPC/PDMS block touches the target substrate first, due to the slide being tilted by a few degrees in the micromanipulator. Full contact between these two surfaces is made controllably in all cases, limiting the speed of the contact area front to less than 1 μm s$^{-1}$ by controlling the z-height of the stage.

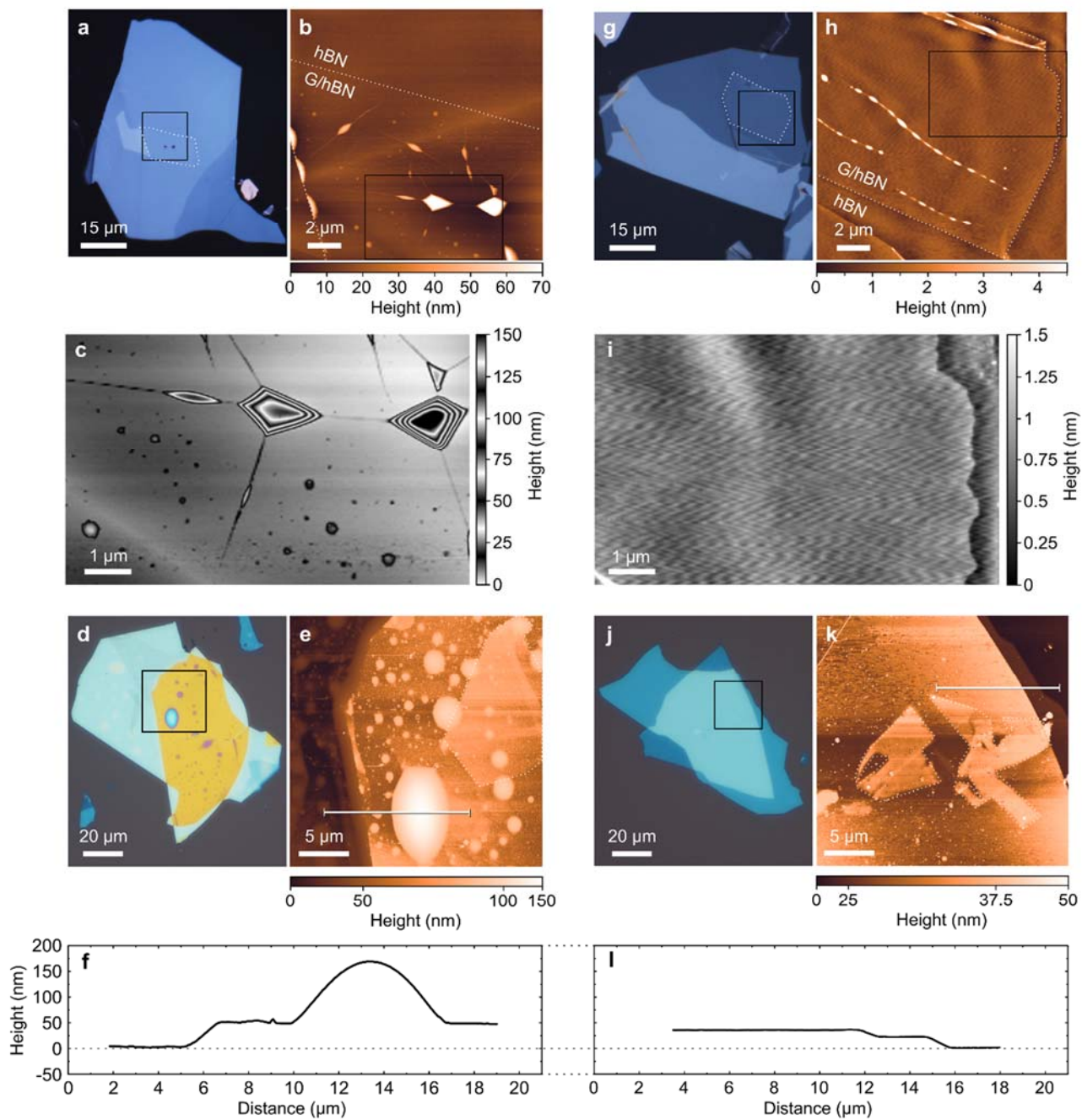

*Figure 3: Atomic force microscopy characterisation of blisters. a Optical image of graphene/hBN stack on PPC (after step iv in Figure 1 f), picked up from pristine SiO$_2$ at 40°C. b AFM topography image of region indicated in (a) – blisters are visible across the stacked region in AFM with heights of ≈ 50 nm. Inset with altered colour scale to show height variation. c Optical image of G/hBN stack dropped down onto a second hBN flake after approaching at 40°C on SiO$_2$ (step vii in Figure 1 f). Many hemispherical blisters are visible over the stack surface. d AFM topography image. Encapsulated graphene flake is visible, indicated by dotted line. e topography line profile of blister from indicated region in (d). f as (a), but with pick-up performed at 110°C. g AFM image shows smaller overall height variation and large blister-free regions. h as (c), but with drop-down performed after approaching at 110°C. No trapped blisters*

*are visible optically or **i** in the AFM scanned region. Encapsulated few-layer graphene regions are indicated. **j** AFM line profile from (i) showing uniform height across the stack. Note that adaptive colour scales are used in (d) and (i) to emphasize topographical features.*

Stacking at 40°C results in apparently blister-free heterostructures, but upon heating above 70°C these blisters become visible and mobile, increase in size, agglomerate and then stabilize within a few seconds (Supplementary Information Movie 1, 2 and 3). These blisters are also afterwards apparent between the graphene and hBN (Figure 3 a-c), and in the finished fully encapsulated graphene sample (Figure 3 d-f). The blisters are up to 100 nm tall and 10 µm across, and evenly distributed over the sample, and account for an area fraction of 20% or more of the total stack area (Figure 3 e). In contrast, assembling at the higher temperature of 110°C results in a complete absence of such blisters (Figure 3 g-l), leading to van der Waals heterostructure devices with greatly reduced interfacial contamination. The surface height variation of the stack is much reduced (Figure 3 i), and no blisters of contamination are visible optically or by AFM (Figure 3 h,i,k,l). We note that the encapsulated graphene flakes are visible in Figure 3 e and k – any trapped interfacial contamination in the stack must therefore either consist of a continuous and homogeneous layer, or be entirely absent.

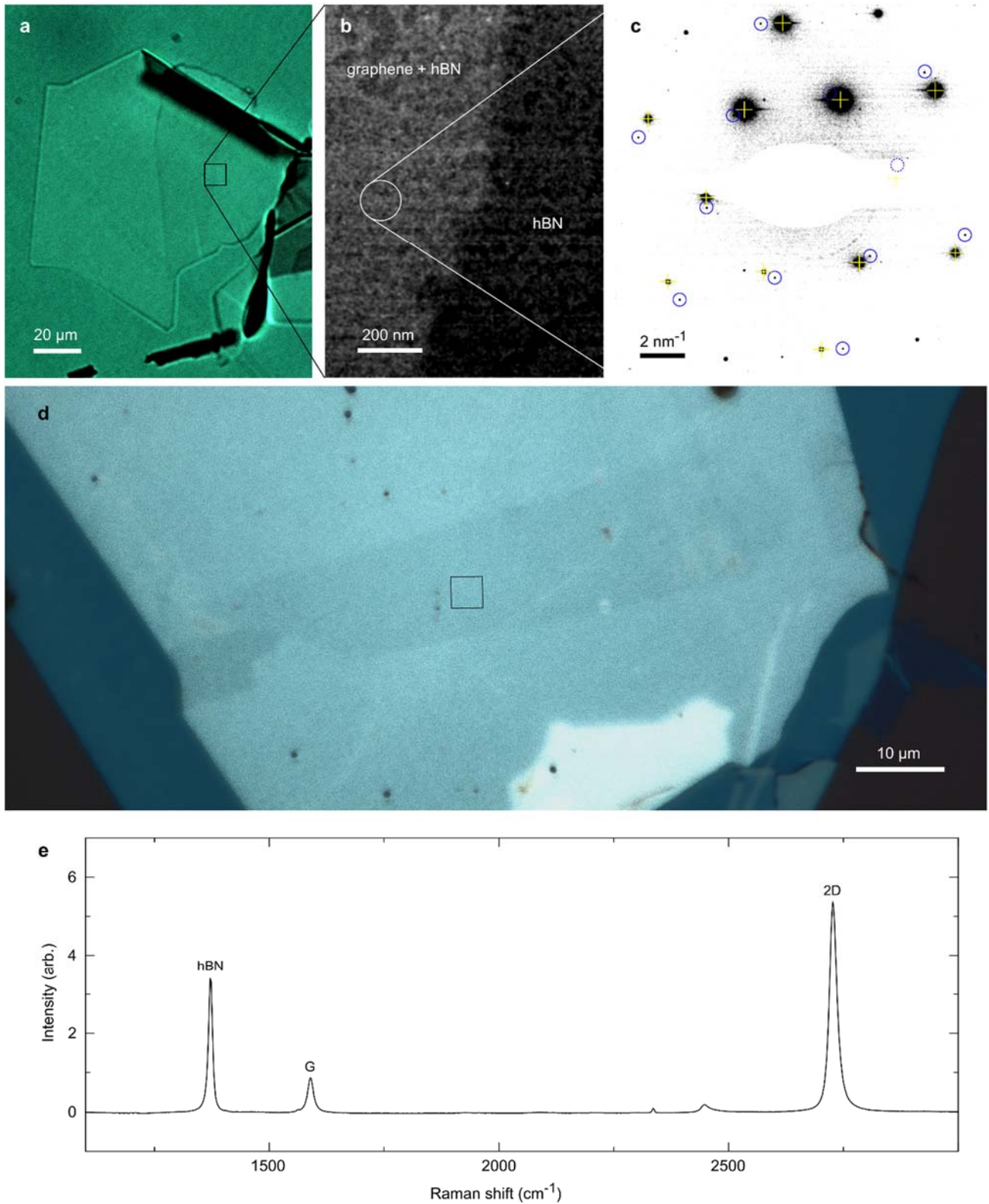

*Figure 4: Transmission electron microscopy and Raman spectroscopy. **a** Optical transmission image of a graphene flake adhered to hBN on PPC/PDMS. **b** Dark-field tilted beam transmission electron microscopy image of region indicated in a. A discontinuous layer of amorphous carbon is visible. The objective aperture has been positioned to provide lighter contrast from the graphene region. **c** Selected area electron diffraction pattern from region indicated in b. The contrast has been enhanced to show both the monolayer*
9

*graphene diffraction pattern (reflections indicated with blue circles) alongside the hBN diffraction pattern (indicated with yellow crosshairs). Points masked by the beam stop or by nearby stronger reflections are indicated with dashed versions of the same.* **d** *Enhanced contrast optical image of the encapsulated graphene. The boxed region indicates the region where the spectrum in (e) was acquired. Maps of the I(2D)/I(G) ratio and G peak G peak full width at half maximum (FWHM) are available in Supplementary Figure 6.* **e** *Raman spectrum of hBN encapsulated graphene fabricated by the hot pick-up method, normalised to the G peak intensity. The I(2D)/I(G) peak ratio is 5.8 (with Lorentzian fits to the peaks [$R^2 > 0.994$] indicating a ratio of 6.12±0.02), with a FWHM of 14.2 ± 0.5 cm$^{-1}$.*

In order to attempt to detect the presence of trapped contamination within stacks, we transferred graphene adhered to hBN to silicon nitride transmission electron microscopy (TEM) aperture grids using the same drop-down methods described above. An optical image of the hBN-graphene heterostructure adhered to PPC/PDMS is shown in Figure 4 a. We then performed tilted beam dark field TEM imaging of the samples by selecting one of the first order graphene reflections. This causes the graphene and any materials with a similar lattice spacing to appear bright in the image. Figure 4 b shows enhanced contrast of graphene with respect to hBN, in addition to an incomplete layer of amorphous carbon which is usually present on 2D materials,[25, 26] and commonly observed in TEM experiments. While we are unable to strictly exclude the possibility that this amorphous carbon is trapped between the layers, we note that it is present everywhere in the dark field image in Figure 4 b, and has not segregated into blisters (which would cause enhanced dark field contrast) as has been observed previously in van der Waals heterostructures. Figure 4 c shows a selected area diffraction pattern of the region indicated in Figure 4 b, where both hBN and monolayer graphene reflections are visible. The lattices have a relative rotation of around 7.5° for this sample, showing that the hBN and graphene lattices have not displayed the self-rotation or 'snap-in' behaviour recently reported in hBN encapsulated graphene heterostructures.[21]

Raman data for a representative hBN encapsulated graphene sample in Figure 4 d is shown in Figure 4 e, with an enhanced contrast optical image of the heterostructure inset showing the collection region. We find a consistent *I(2D)/I(G)* ratio of greater than 5, approaching 6, as previously shown for hBN encapsulated graphene heterostructures, and a *G* peak full width at half maximum (FWHM) of 14±0.5 cm$^{-1}$ over 5-10 μm areas of the heterostructures. Raman maps showing the spatial variation of these values are provided in the Supplementary Information.

*Carrier mobility measurements*

We batch fabricate hBN encapsulated graphene field effect devices with edge contacts in both Hall bar and van der Pauw geometries from blister-free van der Waals heterostructures stacks via electron beam lithography, plasma etching and metal deposition (Figure 5 a). $SF_6$ plasma is used to etch, which we find shows significantly improved selectivity for hBN relative to PMMA (> 45:1), graphene (> 90:1) and $SiO_2$ (> 90:1), rather than the more commonly used $CHF_3$ etch.[10, 11]

Gated pairwise direct current two-point measurements of 284 contacts gives 251 working connections with contact resistances of a few kΩ·μm at the charge neutrality point (CNP), decreasing to a few hundreds of Ω·μm away from the CNP (Supplementary Figure 3), with residual carrier densities less than $10^{12}$ cm$^{-2}$ at zero gate voltage for all monolayer, bilayer and trilayer graphene devices (Figure 5 b, c and Supplementary Figure 4). All our measurements are performed at room temperature with a constant source-drain bias of 5 mV. The total yield of contacts - over 88% - is to our knowledge the largest reported to date in van der Waals heterostructures devices considering the number of contacts produced.

As a result, we are able to present a statistical ensemble of electrical measurements for encapsulated graphene Hall bar devices. Figure 6 a shows room temperature electron and hole mobilities extracted from 55 transconductance measurements of hBN-encapsulated mono-, bi- and trilayer graphene devices. The values of the mobility for bi- and trilayer samples are extracted at saturation, while at the local maximum (induced carrier density of approximately 0.5 - 1 · $10^{12}$ cm$^{-2}$) for single layer devices (see Figure 5 b). For single layer devices, nominal mobilities of up to 117,000 cm$^2$V$^{-1}$s$^{-1}$ were measured, with averages of 44,500 ± 26000 cm$^2$V$^{-1}$s$^{-1}$ and 42,000 ± 24,000 cm$^2$V$^{-1}$s$^{-1}$ for holes and electrons, respectively. All of the field effect mobility values for any pair of monolayer device contacts devices are above 10,000 cm$^2$V$^{-1}$s$^{-1}$, with more than 86% above 20,000 cm$^2$V$^{-1}$s$^{-1}$.

In the diffusive limit the mean free path $\lambda_{mfp}$ can be found from the semiclassical conductivity[22, 23] $\sigma = (2e^2/h)k_F \lambda_{mfp}$, where $k_F = \sqrt{\pi n}$ is the Fermi wave number, $n$ is the carrier density, $e$ is the elemental charge and $h$ is Planck's constant. In our calculations of mobility, we assume that scattering is dominated by charged impurities. The high mobilities we calculate indicate that in the majority of cases in our monolayer devices the transport is limited by boundary scattering since the calculated mean free path is comparable to or larger than $w$, the device width[20] - in particular 70% of devices show $\lambda_{mfp} \geq w/2$ (Figure 6 b, d). Furthermore 16% of the measurements are strictly quasiballistic having the estimated mean free paths exceeding the device width.[27, 28]

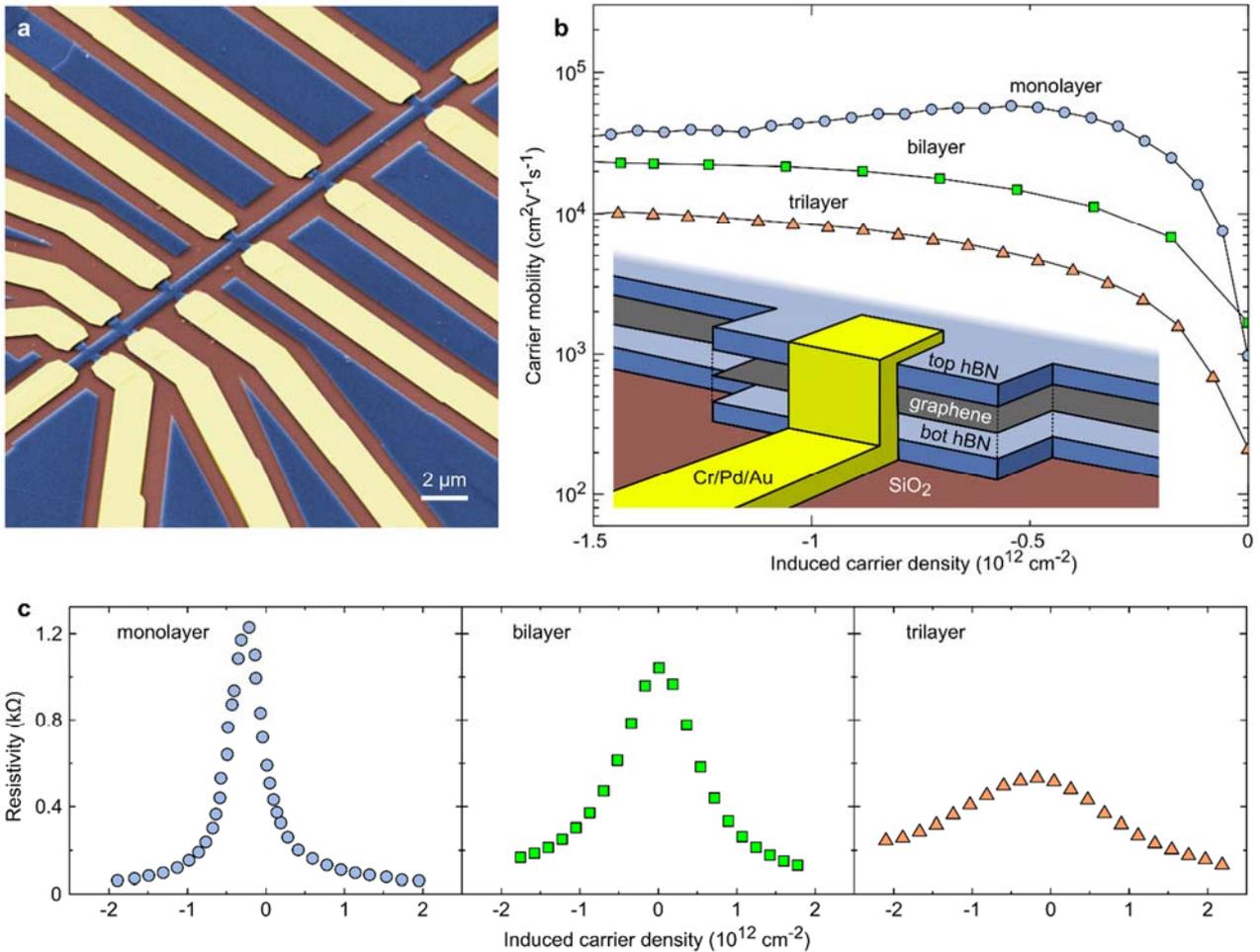

*Figure 5: Electrical measurements of Hall bar devices. a False colour scanning electron micrograph of encapsulated graphene device. Blue regions are unetched hBN and hBN encapsulated graphene, brown are exposed $SiO_2$, and yellow regions are metal contacts. b Room temperature field effect mobility measurements for representative monolayer (#4), bilayer(#11) and trilayer (#14)*

*Hall bar devices. A monolayer device displaying diffusive transport is shown. Blue, green and orange indicate monolayer, bilayer and trilayer results respectively. Inset shows exploded schematic of contact region. **c** Resistivity vs. gate-induced carrier density for monolayer, bilayer and trilayer graphene Hall bar devices. Color scheme as in (b).*

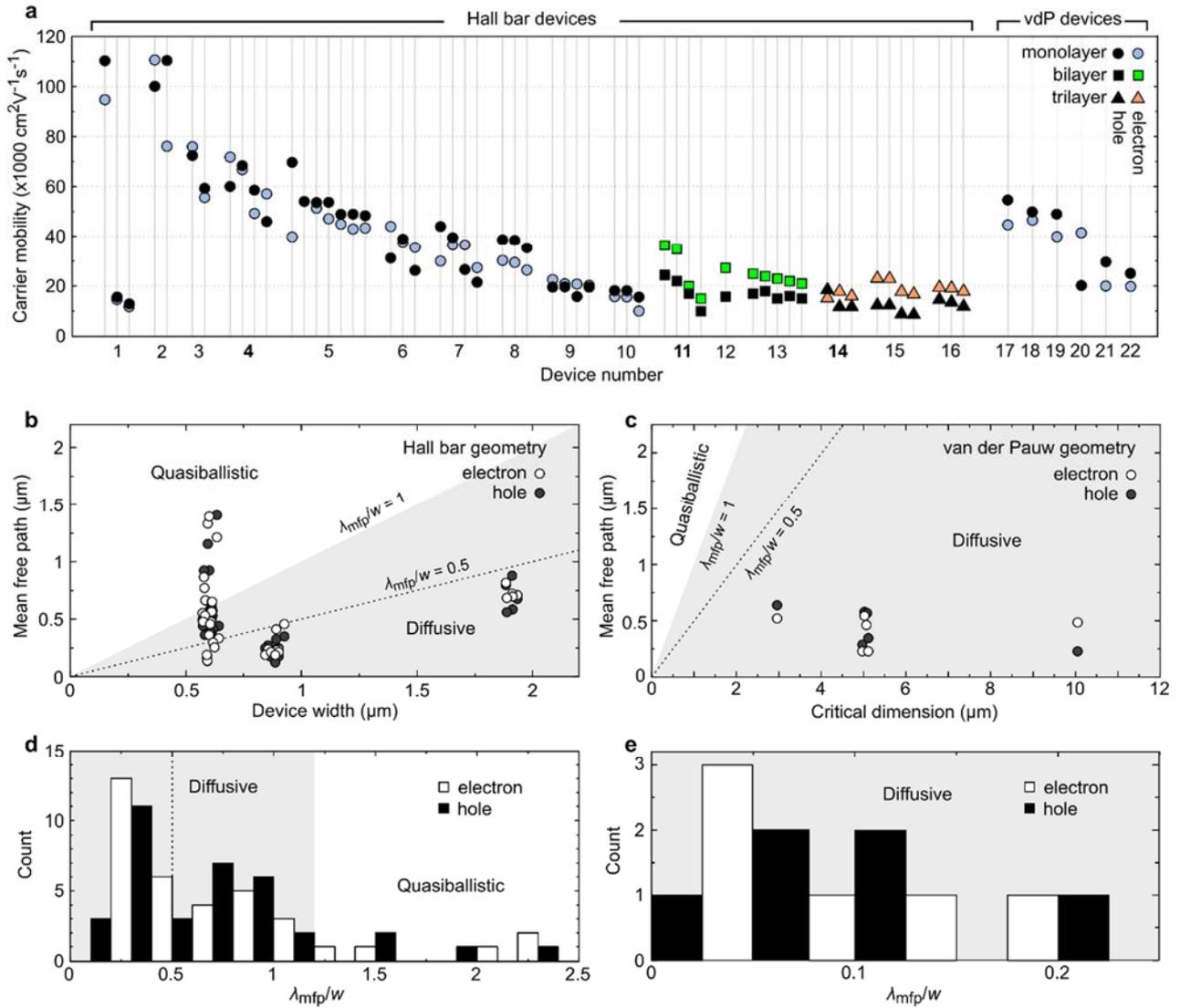

*Figure 6: Mobility and mean free path statistics. **a** Room temperature field effect carrier mobility plot for all the devices (Hall bar and van der Pauw) in this study. **b** Plot of electron and hole mobility for all monolayer Hall bar devices. Quasiballistic ($\lambda_{mfp} \geq w$) and diffusive ($w > \lambda_{mfp}$) regimes are indicated by the gray region. Note that 60% of our narrow channel devices have a calculated mean free path larger than w/2 (dashed line), a clear indication of transport affected by boundary scattering.[17] **c** Plot of electron and hole mobility for all monolayer van der Pauw devices. Quasiballistic ($\lambda_{mfp} \geq w$) and diffusive ($w > \lambda_{mfp}$) regimes are indicated by the gray region. The average mean free path is 450±170 nm and 420±140 nm for holes and electrons respectively. **d** Histogram of monolayer electron and hole mobilities for all Hall bar devices. **e** Histogram of monolayer electron and hole mobilities for all van der Pauw devices.*

Three bilayer and three trilayer graphene samples produced in the same way display field effect mobilities as high as 37,000 cm$^2$V$^{-1}$s$^{-1}$ and 23,000 cm$^2$V$^{-1}$s$^{-1}$, respectively. The mean field effect hole and electron mobilities are 17,000±4,000 cm$^2$V$^{-1}$s$^{-1}$ and 24,500±7,000 cm$^2$V$^{-1}$s$^{-1}$ for the bilayer devices, and 13,000±3,000 cm$^2$V$^{-1}$s$^{-1}$ and 19,000±2,500 cm$^2$V$^{-1}$s$^{-1}$ for the trilayer devices, respectively.

In addition to the Hall bar devices we fabricated six encapsulated monolayer devices in van der Pauw (square) geometries of 3, 5 and 10 μm side length with four contacts placed at the corners of the device edges. For such devices we find hole and electron mobilities of 38,000±15,000 cm$^2$V$^{-1}$s$^{-1}$ and 36,000±12,000 cm$^2$V$^{-1}$s$^{-1}$ respectively. The average mean free path for carriers in all such devices is 450±170 nm and 420±140 nm at room temperature. A representative gate sweep is available in the Supplementary Figure 7 while the mean free path of sample 2 is shown to reach 0.9 μm at a carrier density of 10$^{12}$ cm$^{-2}$ in Supplementary Figure 8.

In order to prove the versatility of our technique, top-gated samples were fabricated by the drop-down of an additional hBN flake to cover the entire device area (Supplementary Figure 5). This flake prevents short-circuiting during subsequent deposition of a metallic top-gate and contacts. The transconductance of the devices was compared (Supplementary Figure 5) suggesting that top gates fabricated by drop down are comparable in performance and quality to hBN and oxidised silicon bottom gates, which we attribute to the absence of trapped contamination.

**Discussion**

The presented method for stacking 2D materials allows the preparation of clean heterostructures with a high throughput due to a higher yield of large exfoliated single layer crystals, the parallelization of the different fabrication steps enabling batch production and with close to 100% pick-up yield.

The fact that the observed reduction of *I(2D)/I(G)* peak ratio in graphene flakes micromechanically cleaved on plasma treated oxidised silicon is fully reversed upon subsequent pick-up with hBN shows that oxygen plasma treatment of the substrate does not introduce point defects or charged contaminants attached to the exfoliated graphene. As a result, oxygen plasma pre-treatment is a suitable means of producing large numbers of appropriately sized exfoliated 2D materials on oxidised silicon for the fabrication of prototype van der Waals heterostructures, which consistently exhibit clean interfaces and high carrier mobilities.

The complete elimination of blisters of contamination formed between flakes assembled at 110°C versus the large number and size of blisters formed between flakes assembled at 40°C or less, suggests that at low temperatures the surfaces of the flakes are covered by largely immobile adsorbants, which become more mobile at higher temperatures. This leads to the observed segregation of this layer into blisters during stacking at 40°C and upon subsequent heating above 70°C. We find it extremely difficult to pick up crystals cleaved on plasma-treated substrates at 40°C, despite an extended contact time of several hours, indicating that the adsorbants are effectively immobile in this case and interfere critically with the adhesion between flakes. The higher adsorbant mobility at 110°C on the graphene surface allows them to diffuse across the surface of the 2D materials, away from the van der Waals contact front which proceeds in a controlled way across the surface of the flakes during assembly, and appears to gather at the edges of flakes (Supplementary Movie 4).[22] We infrequently observe inconsistent behaviour during drop down in some samples (Supplementary Movie 5) that can lead to variation in the Raman *I(2D)/I(G)* ratio and *G* peak full width at half maximum of the encapsulated graphene in limited areas of the surface (Supplementary Figure 6) – this highlights the importance of control of the drop down step in making good contact between the layers of vdW heterostructures, here minimising doping variations. Reduction in the viscosity of PPC at higher temperature may also allow a more conformal contact to be made between the 2D materials during stacking. It is interesting to note that the adhesion force between graphene and hBN resulting from stacking at temperatures above 110°C is sufficient to overcome the adhesion of graphene to silicon oxide, but that adhesion between the PPC and the exposed graphene outside the hBN area is not strong enough to delaminate the graphene from the oxidised silicon. This points to the gradual exclusion of adsorbants or contaminants at the contact front between two 2D materials as the determining factor for optimisation of adhesion forces and for establishing a good van der Waals contact. This also raises the question of what materials the contaminants and adsorbants consist of. Since the process is performed in ambient atmosphere the three most likely sources of contamination are adsorbed water and/or trapped air, deposited ambient airborne hydrocarbon contamination[25] or residues from the tape used for micromechanical cleavage produced during thermal release of the flakes on the substrate. The temperature of 110°C needed to render these contaminants sufficiently mobile to be excluded from the advancing van der Waals contact interface suggests adsorbed water as the most likely candidate. This would account for the observation of blister formation from apparently clean vdW heterostructures during heating to 70°C – the adsorbed water coalesces as the vdW interface area increases, and vaporization of water causes the large volume blisters to be produced.

Atomic force microscopy shows that any residual contamination between layers in bubble free regions must be either uniform or completely absent, since we are able to detect changes in thickness of the graphene encapsulated in hBN even down to a single monolayer (Figure 3 i and k). Dark field and selected area diffraction patterns from transmission electron microscopy studies (Figure 4 b and c) provide further evidence that the interface between the hBN and graphene does not include any contamination, along with providing evidence that the pick-up technique as described here does not result in self-aligning rotation of 2D materials.[21] Finally the Raman spectroscopy of graphene encapsulated in hBN (Figure 4 d, e), and Supplementary Figure 6) shows a *I(2D)/I(G)* ratio approaching 6 and a *G* peak FWHM of 14 cm$^{-1}$, which serve as evidence of a lack of doping of the graphene within the stack – results which are also confirmed by our electrical measurements of the low doping of our samples: all show residual carrier densities greater than $10^{12}$ cm$^{-2}$ at zero gate bias.

The presented techniques also enable us to pick up pre-patterned 2D materials, despite the presence of PMMA residues on top of the patterned flakes, an inevitable by-product of electron beam lithography. This means both that PMMA residues are largely excluded from the surface via this technique, and also that it is not a prerequisite of this technique to have atomically flat and clean surfaces before carrying out the stacking procedure, as presented previously.[10] This removes the limitation of simultaneous, through-stack lithography as the only option for the construction of van der Waals heterostructures devices, allowing more complex and flexible architectures to be realised by enabling the clean stacking of separate, differently shaped 2D materials analogous to modern integrated circuits with two or more layers of active electronic components or interconnects in a single circuit.[29]

The devices produced using this hot stacking technique display very high room temperature mobilities, with 55% of the measurements in monolayer devices exhibiting mean free paths larger than the half of the device width as well as uniformly low contact resistance. This high consistency is achieved without any annealing[10, 12, 16, 21] or further cleaning step of the stacks.

While little data is available in literature concerning the yield of heterostructure device fabrication, we believe that our stacking yield of nearly 100% and contact fabrication yield of 88% represent significant progress towards consistent, reproducible device fabrication of van der Waals heterostructures. We find that statistically significant figures for these yields is unfeasible to obtain without batch fabrication, and that the high yield across many samples is a direct consequence of the complete exclusion of interfacial contamination during stacking at high temperatures. In comparison, stacking at lower temperatures lead to blisters occurring over an interface area fraction of 20% or more which strongly decreases the fabrication yield.

The mean carrier mobilities measured in bilayer and trilayer graphene devices of 21,000±7,000 $cm^2V^{-1}s^{-1}$ and 16,000±4,000 $cm^2V^{-1}s^{-1}$, with peak values of 37,000 $cm^2V^{-1}s^{-1}$ and 23,000 $cm^2V^{-1}s^{-1}$, are consistent with the highest values reported at room temperature in the literature for bilayers[31] and are the highest reported for trilayers at room temperature.[32]

We note that in all but one instance, the electron mobility exceeds the hole mobility for bilayer and trilayer samples (Figure 6 a). This trend is largely reversed in monolayer samples, and is consistent with published observations.[33]

In 16 % of the measurements the calculated room temperature mean free path exceeds the sample width[10, 20, 34] (see Figure 6 b, d). In this limit, however, the semiclassical transport model used to estimate the mean free path is no longer strictly valid, implying that both the highest values of mean free path and the carrier mobility should be regarded with caution.[20] We speculate that the higher mobility and thus stronger tendency of ballistic transport behaviour for the monolayer samples is responsible for the larger spread in carrier mobility as compared to bi- and trilayers (Figure 6 a). The spread in carrier mobility is a consequence of transport in ballistic samples being more sensitive to boundary effects such as possible edge disorder and the exact device geometry.[10, 20, 27] The mean free paths for the van der Pauw geometry samples (around 0.5 µm) are consistent with the onset of large variation in the calculated mobilities for the narrowest channel Hall bar samples (Figure 6 b-e), which supports the conclusion that boundary effects have a strong influence at such scales and lead to the observed large variations in the calculated mobilities here.

In conclusion, we have presented a facile and robust technique for the batch fabrication of van der Waals heterostructures, demonstrated by the controlled production of 23 mono-, bi- and trilayer encapsulated graphene devices. Stacking at elevated temperatures (higher than 110°C) results in high mean carrier mobilities for bilayer and trilayer samples of 21,000±7,000 $cm^2V^{-1}s^{-1}$ and 16,000±4,000 $cm^2V^{-1}s^{-1}$. The majority of monolayer Hall-bar devices exhibit transport limited by the edges, with mean-free paths exceeding half of the channel width. This stacking technique enables the pick-up and drop-down of flakes of 2D materials at desired locations with near 100% yield, and with a yield of 88% for subsequently fabricated electrical contacts with uniformly low contact resistance.

The absence of trapped contamination in our samples, manifesting as blisters between the stacked flakes, indicates that adsorbants are largely or completely excluded from between the flakes during the drop-down procedure, being pushed out in front of the proceeding van der Waals contact region. In addition, the high temperature used for drop down may also allow a more conformal contact between 2D materials during stacking as a result of reduction of viscosity of the PPC. The presented method readily lends itself to fabrication of any desired van der Waals heterostructures, completely avoiding contact with liquids, whether in ambient conditions or in controlled atmospheres. The technique even permits the pick-up of lithographically patterned 2D materials and integration into van der Waals heterostructures, or for additional layers to be dropped down onto pre-existing devices. By picking up device layers such as graphene with encapsulating layers such as hBN, the need for multiple separate encapsulation steps is avoided allowing efficient batch fabrication of heterostructure devices.

The ability to produce a large number of devices with high yield is a key advantage of this process, and paves the way for the statistical studies of device performance. Such studies are essential to gain an understanding of van der Waals heterostructure based device performance in a technological perspective, for fundamental research and for further progress towards real-world device applications of 2D materials.

**Methods**

*Cleaving of graphene and hBN*

Graphene and hBN are cleaved on 100 nm Si oxide thermally grown on standard 4 inch Si wafers. Natural graphite crystals (NaturGrafit GmbH), and hBN bulk crystals (HQgraphene) were mechanically exfoliated with Nitto Denko SWT 20+ die sawing tape. Oxidised silicon is treated in oxygen plasma for 3 minutes (PlasmaEtch PE-50, 300 mbar $O_2$, 120 W), and the 2D material loaded tape is immediately applied to the silicon oxide surface. The tape is subsequently released from the surface by heating to 85°C on a hot plate.

*Preparation of the glass slide for pick-up and drop-down*

The PDMS is prepared from SYLGARD® 184 by mixing 10 parts base and 1 part curing agent and cured at 70°C overnight, leading to a 1 mm thick layer. The PDMS is treated with oxygen plasma as above for 10 min and then a PPC layer is spun on top of it (15% in anisole, 50K, 1500 rpm), resulting in a thickness of 5 μm. Double-sided tape is used to attach a 1 x 1 mm$^2$) piece of PPC coated PDMS to a glass microscope slide.

*Raman Spectroscopy*

Raman spectra are taken with a Thermo Fisher DXR Raman spectrometer using a 455 nm laser source with a power of 1mW and 20 s duration multiple exposures. Raman spectra and maps for encapsulated graphene were acquired in a Thermo Fisher DXRxi Raman spectrometer using a 455 nm source, with a power of 10 mW and 20 s duration to provide adequate signal to noise ratio from the graphene within the heterostructure.

*AFM*

The AFM scans are done in a NTEGRA scanning probe microscope from NT-MDT with a Smena measuring head. The scans are performed in tapping mode with typical parameters of a driven frequency of 340 Hz, a magnitude of 10 nA, a set point of 5 nA and scan speed of 10-30 µm s$^{-1}$.

*TEM*

Tilted beam dark field images were acquired in a Tecnai T20 G2 operated at 200 kV by selecting a first order graphene reflection with an objective aperture with a diameter of 2 nm$^{-1}$ in the diffraction plane. Selected area aperture diffraction patterns were taken in the same instrument with a 200 nm diameter SAD aperture.

*Device Fabrication*

The EBL is performed in a JEOL JBX-9500FS, with an acceleration voltage of 100 kV. Poly(methyl methacrylate) (PMMA) is used as resist in the EBL process, a 4 wt% PMMA 996K in anisole solution is spun on the chips (1 min at 1500 rpm, acceleration 500 rpm/s), followed by a post-bake (10 min at 150°C).

*Etching*

The etching of the stack is performed in a SPTS ICP Etch, $O_2$ is used to etch the graphene and $SF_6$ is used to etch the top and bottom hBN. Metal contacts of 2 nm Cr, 15 nm Pd and 30 nm Au are deposited by Physimeca ΦSES250 electron-beam evaporation with low rates (approximately 1 Å s$^{-1}$ for Cr and Pd and approximately 3 Å s$^{-1}$ for Au).

The authors declare that all data supporting the findings of this study are available within the article and its supplementary information files.

**Acknowledgements**

This work was supported by the Danish National Research Foundation Center for Nanostructured Graphene, project DNRF103, the EU Seventh Framework Programme (FP7/2007-2013) under grant agreement number FP7-6040007 "GLADIATOR" and the EC Graphene FET Flagship, grant agreement number 604391. The A.P. Møller and Chastine Mc-Kinney Møller Foundation is acknowledged for their contribution toward the establishment of the Center for Electron Nanoscopy at the Technical University of Denmark.

**Contributions**

F. P. conceived the experiments. F. P. and L. G. fabricated the samples. B. S. J. and L. W. contributed in the development of the techniques. F. P., L. G., J. M. C. and T. B. performed the measurements and data analysis. T. B. performed TEM analyses. P. B. and J. H. contributed with data analysis and interpretation. F. P., T. B. and P.B. wrote the manuscript. All authors discussed the results and commented on the manuscript.

**Competing financial interests**

The authors declare no competing financial interests.

**Supplementary Information**

Supplementary Figures 1-14, Supplementary Movie 1-5, Supplementary Method details.

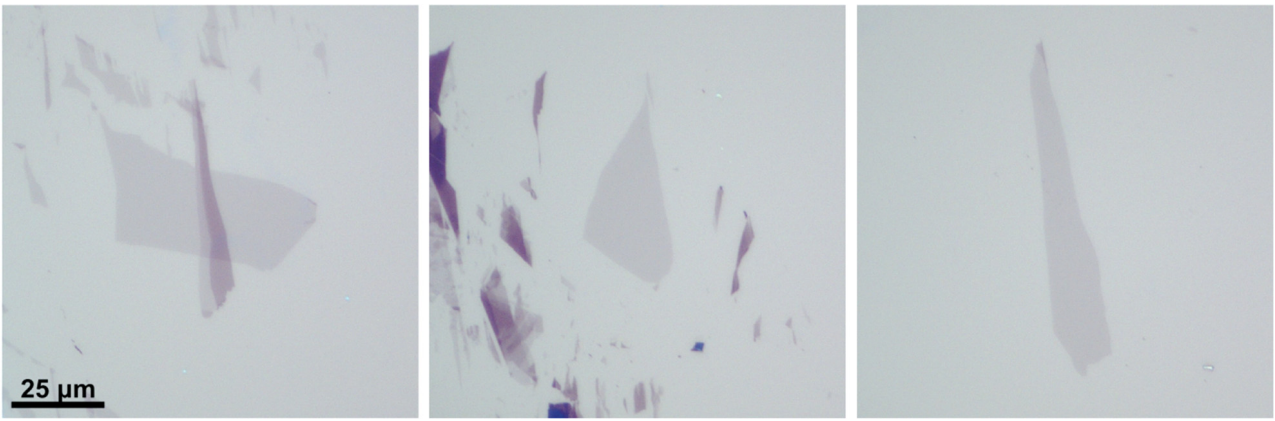

**Supplementary Figure 1: Micromechanical cleavage of graphene on oxygen plasma treated Si/SiO$_2$.** Optical microscopy images of three examples of large single layer graphene flakes cleaved on a single Si/SiO$_2$ chip after oxygen plasma treatment.

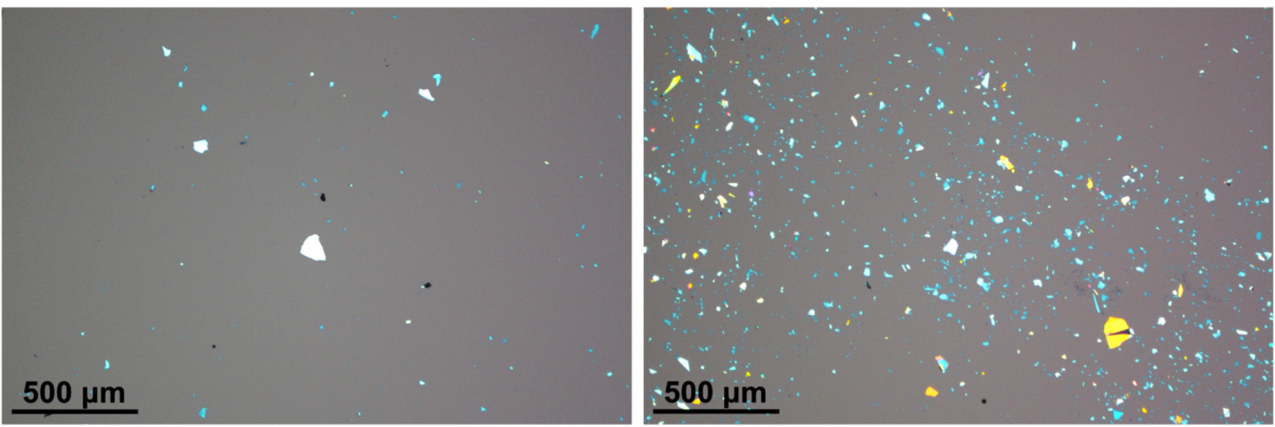

**Supplementary Figure 2: Comparison of hBN yield.** Optical images showing the difference in cleaving yield of hBN on pristine (left) and oxygen plasma-treated (right) SiO$_2$ surfaces, using the same cleaving tape.

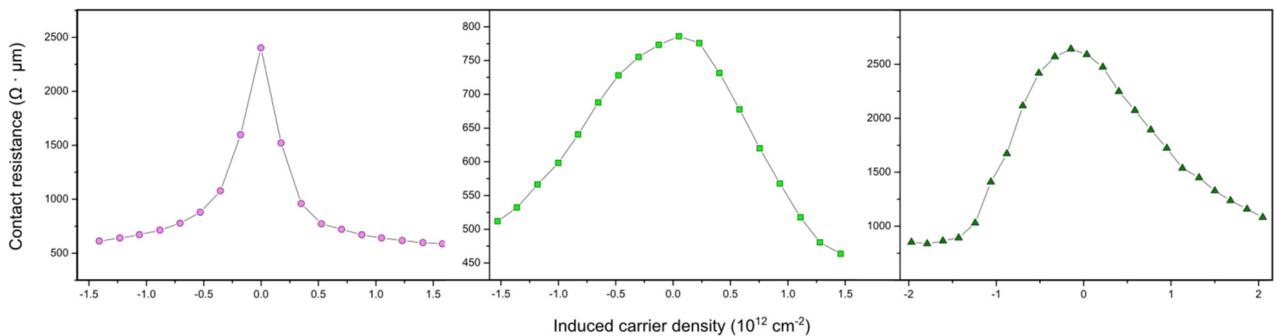

**Supplementary Figure 3: Contact resistance measurements of encapsulated graphene.** Contact resistance measurements from single- (left), bi- (middle) and trilayer (right) graphene devices. The values are extracted from the measurements of source-drain current and $R_{xx}$.

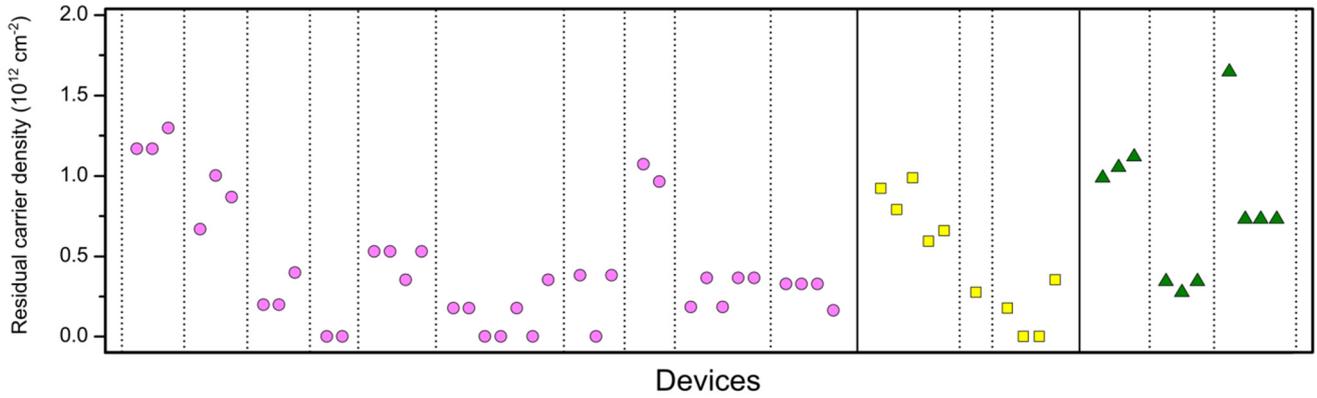

**Supplementary Figure 4: Doping of encapsulated graphene.** Residual carrier density at zero bias for the Hall bar device measurements reported in Figure 6.

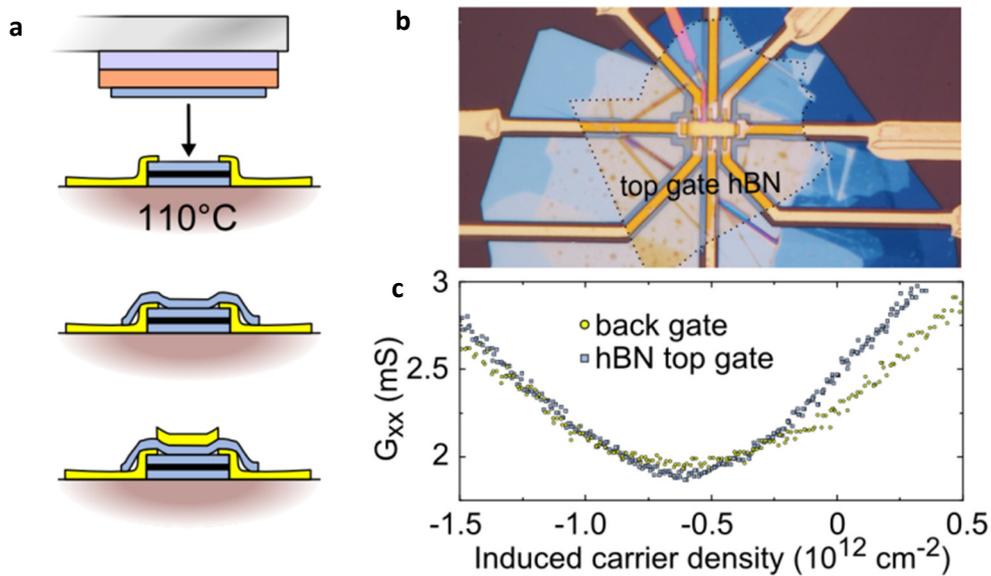

**Supplementary Figure 5: Top gate addition.** a) Schematic process flow for addition of hBN-insulated top gate. b) Optical image of top gate insulator hBN dropped down onto device after fabrication. c) Conductance vs. gate-induced carrier density for the same device using Si back gate and hBN top gate.

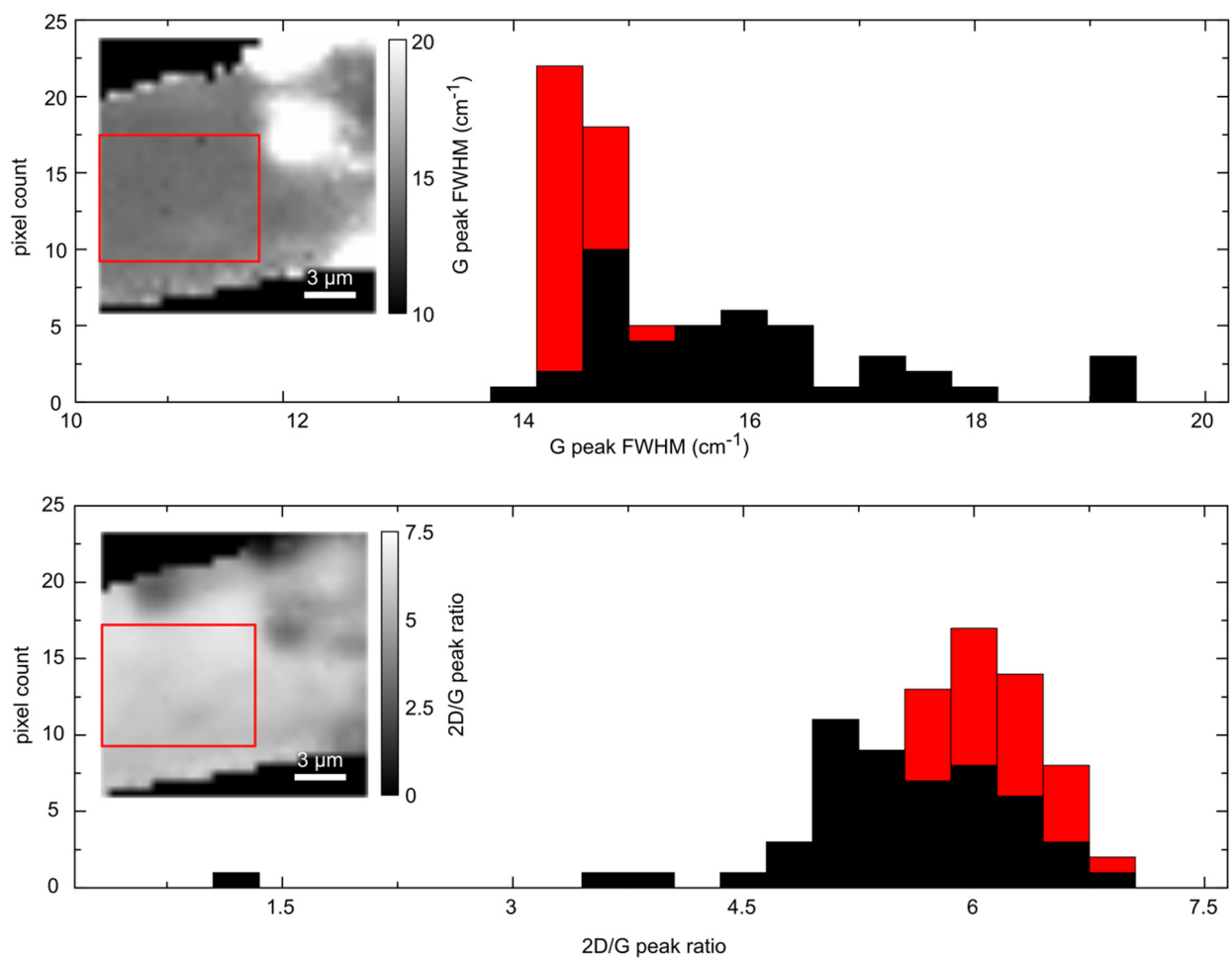

**Supplementary Figure 6: Homogeneity of Raman characteristics.** Raman maps of hBN encapsulated graphene, taken from the indicated region in Figure 4d inset in the main manuscript. Histograms show pixel counts of the measured parameters, the *G* peak full width at half maximum (FWHM) and *I(2D)/I(G)* peak intensity ratio. Contributions to the distributions from the boxed regions in the maps are marked in red in the histograms, whilst the black areas of the histograms represent the values outside of the areas marked in red. These regions are specifically chosen to show variation in the Raman properties of encapsulated graphene dependent on the control of the contact front (see Supplementary Movie 5). The drop down techniques used in this study result in large regions of encapsulated graphene with a narrow distribution of *G* peak FWHM around 14-15 cm$^{-1}$ and an *I(2D)/I(G)* peak intensity ratio of around 6. These values demonstrate the low residual doping of such samples.

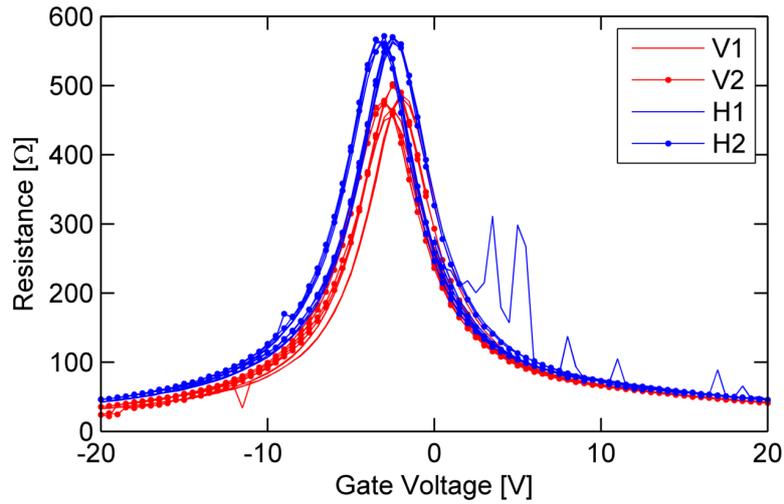

**Supplementary Figure 7: van der Pauw measurements of square devices.** Representative gate voltage sweep for a van der Pauw device with 5 μm side length – device #21 in Figure 6a of the main manuscript. Reproducible gate induced changes in the resistance irrespective of the configuration of the contacts (V1, V2, H1, H2) imply that the device is homogeneous and that the reciprocity theorem holds. See Supplementary Methods.

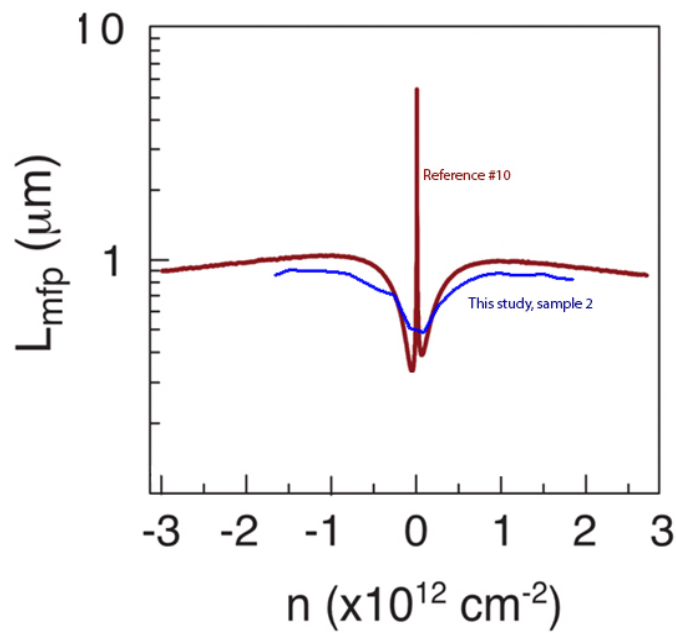

**Supplementary Figure 8: Mean free path comparison.** Comparison of mean free path in this study (blue curve) vs. data from reference 10 (main text references) (brown curve) from a 1.9 μm wide Hall bar device – device #2 in Figure 6a of the main manuscript. The figure is adapted from reference 10 (main text references).

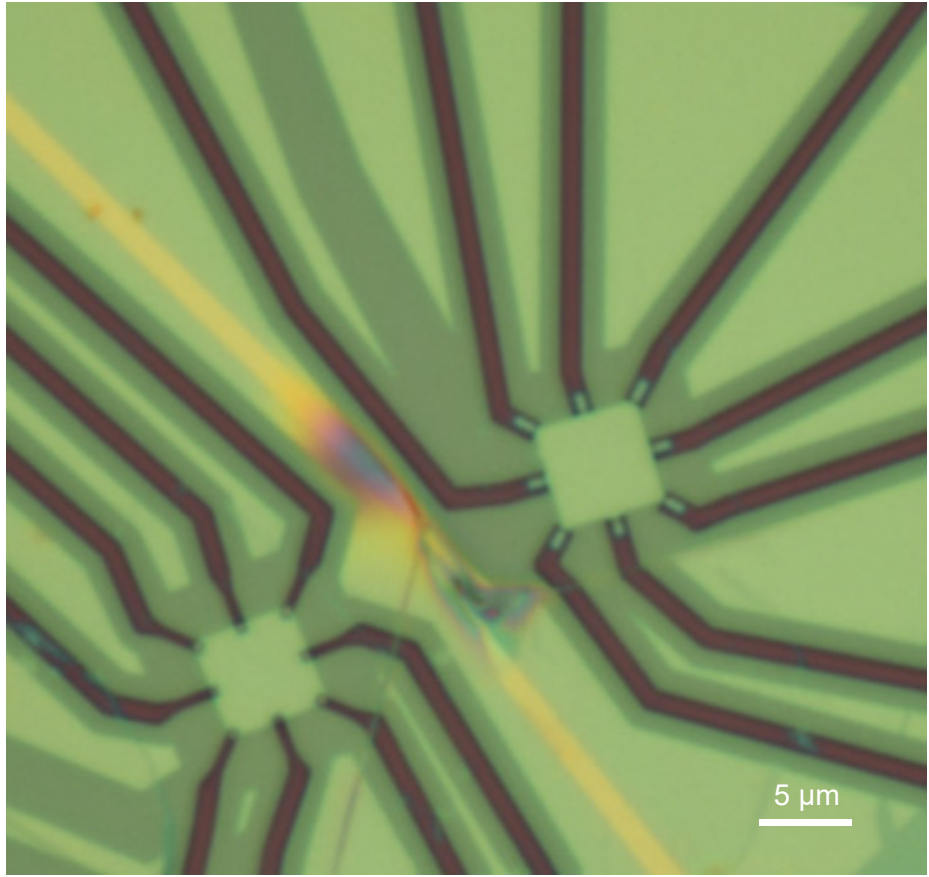

**Supplementary Figure 9: Fabrication of van der Pauw devices.** Optical image of two van der Pauw type encapsulated devices before deposition of metal electrodes and lift-off.

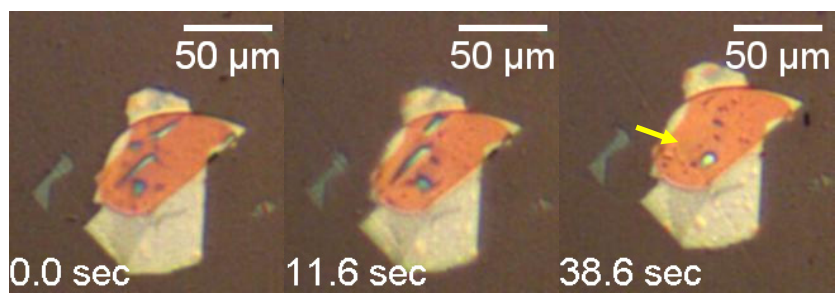

**Supplementary Figure 10: Trapping and migration of blisters in hBN/graphene/hBN.** Optical image of mobile blisters (indicated by yellow arrow) within an hBN-graphene-hBN heterostructure on $SiO_2$ assembled at 40°C and heated to 70°C (selected frames from Supplementary Movie S1).

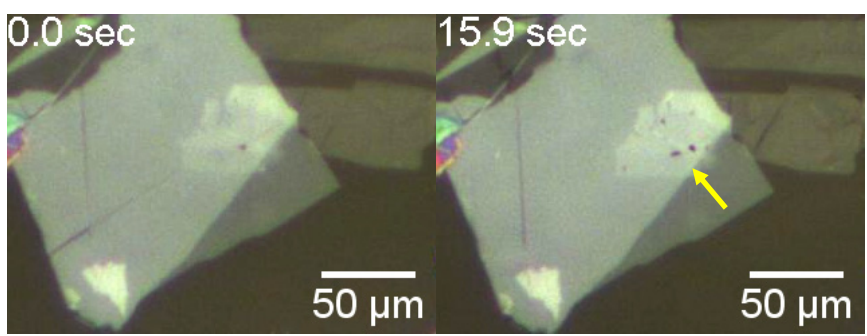

**Supplementary Figure 11: Trapping and migration of blisters after pick-up.** Optical image of mobile blisters (indicated by yellow arrow) within an hBN-graphene heterostructure on PPC/PDMS before drop-down on hBN, assembled at 40°C and heated to 70°C (selected frames from Supplementary Movie 2).

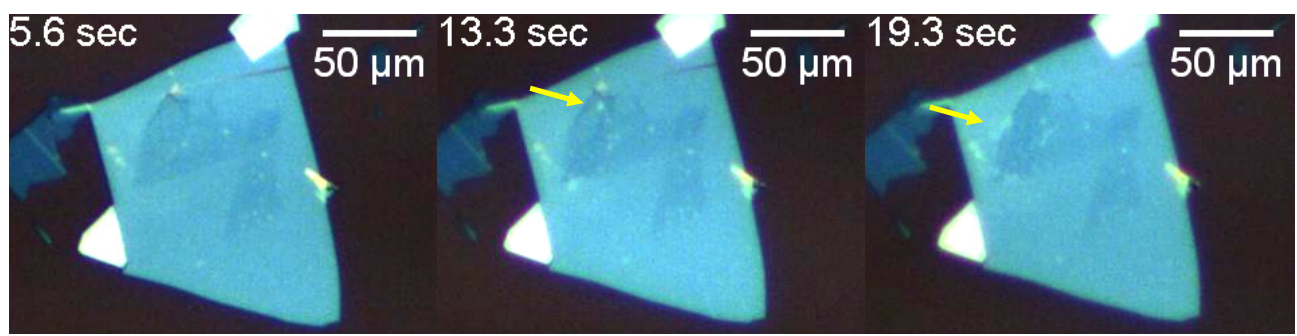

**Supplementary Figure 12. Trapping and migration of blisters after drop-down.** Optical image of mobile blisters (indicated by yellow arrow) within an hBN-graphene heterostructure on $SiO_2$ before pick-up assembled at 40°C and heated to 70°C (selected frames from Supplementary Movie 3).

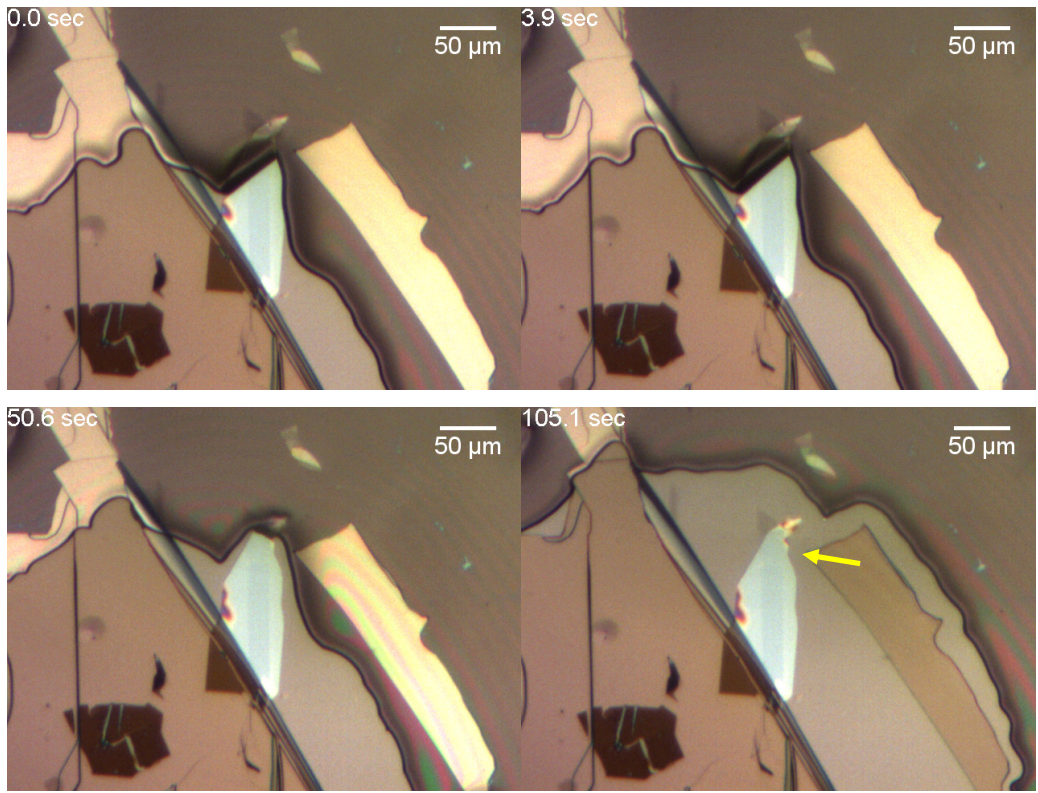

**Supplementary Figure 13: Blister-free drop-down.** Drop-down of an hBN flake adhered to PPC/PDMS onto a graphene layer on $SiO_2$. Control of the contact front forces any contamination between the faces of the 2D materials out of the heterostructure towards the edges (indicated by yellow arrow). Selected frames from Supplementary Movie 4.

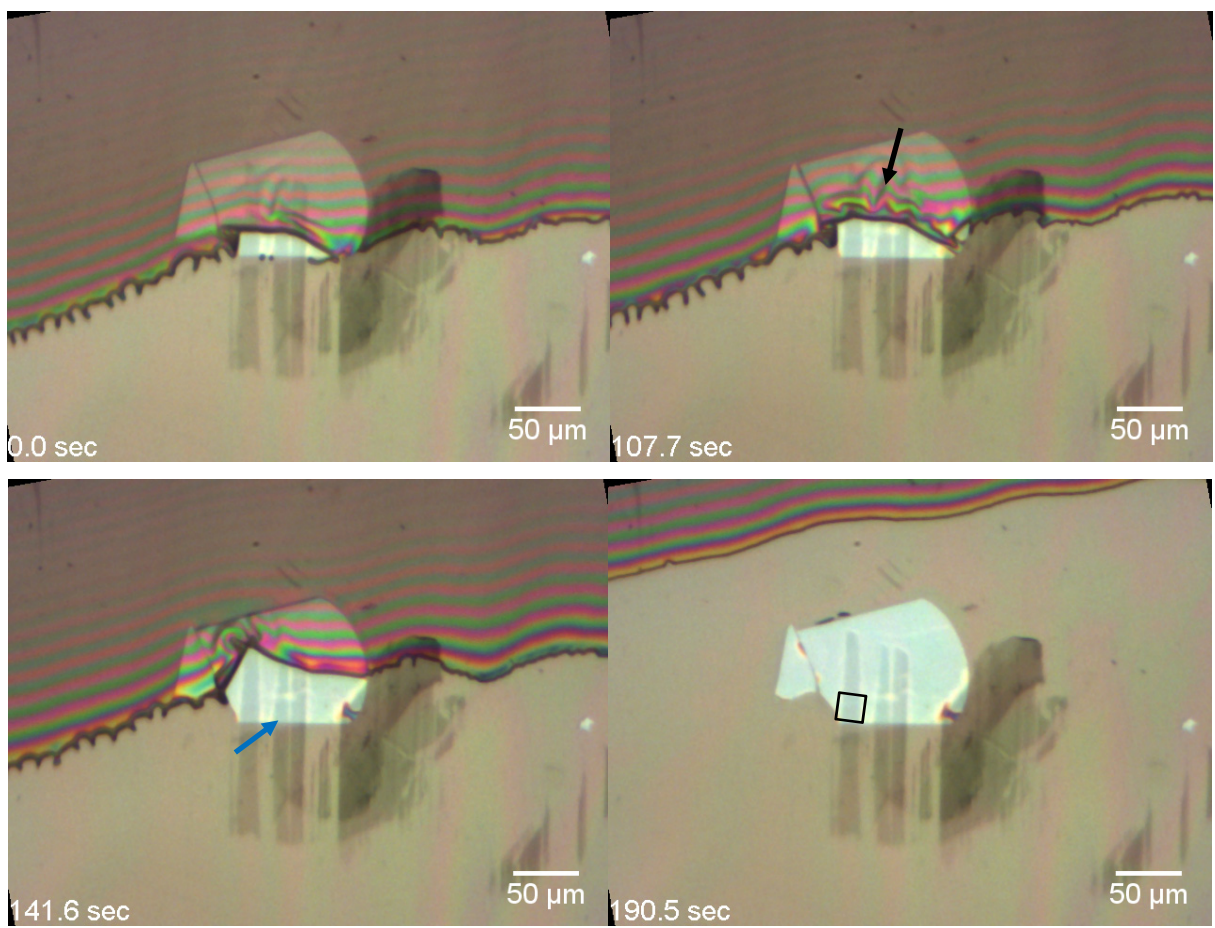

**Supplementary Figure 14: Blister formation during drop-down.** Drop-down of an hBN flake adhered to PPC/PDMS onto a graphene layer on $SiO_2$. In this instance, the contact front is not well controlled (black arrow), resulting in the visible formation of blisters (blue arrow) and inhomogeneous Raman *I(2D)/I(G)* and *G* peak FWHM values (See Supplementary Figure 6). The boxed region indicates where the data in Supplementary Figure 6 was acquired. Selected frames from Supplementary Movie 5)

**Supplementary Methods**

Details of the hot pick-up transfer are described in the following. Step 1: Pick up of the hBN flake from the silicon oxide surface. A piece of $Si/SiO_2$ with exfoliated hBN is placed on a microscope vacuum hot plate stage. The glass slide is held in a micromanipulator and kept approximately horizontal above the sample, with the polymer facing down. The PDMS/PPC is placed in contact with the hBN flake by lowering the glass slide along the *z*-axis of the micromanipulator. When the polymer completely covers the hBN flake, the temperature is raised to 55°C, and then lowered to 40°C. At this temperature, retracting the glass slide results in the pick-up of the hBN from the $SiO_2$ on the PPC (Figure 1 f i). This procedure is repeated for all the targeted hBN flakes (one per slide), so that a batch of hBN flakes adhered to glass slides is prepared.

Step 2: Drop-down of hBN on top of graphene. A Si/SiO$_2$ substrate with exfoliated graphene flakes is placed on the heating stage. The temperature is raised to 110 °C. A previously picked-up hBN flake is aligned over a chosen graphene flake with some separation between the surfaces (Figure 1 f ii). The glass slide is then lowered until the polymer comes into contact with the oxide surface. By further lowering the glass slide, the PPC front proceeds until reaching full contact with the SiO$_2$ surface. In this phase the polymer behaves as a liquid. It is important to proceed as slow as possible, in order to avoid non-conformal contact between hBN and graphene. When fully in contact, by moving slowly the glass slide away from the surface, the polymer front recedes very slowly. In this way it is possible to release the hBN from the PPC onto the graphene flake, without detaching the PPC from the PDMS block (Figure 1 f iii). The procedure is repeated until all the graphene flakes are covered by the hBN flakes prepared in 'Step 1'.

Step 3: baking. In order to promote the adhesion between graphene and hBN, the Si/SiO$_2$ substrate with the stacks is baked on a hot plate (in air) at 130 °C for 15 min.

Step 4: pick-up of the hBN/G stacks. All the hBN/G stacks are picked up from the SiO$_2$ surface repeating the procedure introduced in 'Step 1', one per glass slide (Figure 1 f iv).

Step 5: drop down of the hBN/G stack on the bottom hBN. Finally, the hBN/G stacks (on PPC) are dropped down on hBN flakes (cleaved on Si/SiO$_2$ substrates) similarly to what is done in 'Step 2' (Figure 1 f v-vii). The bottom hBN flakes are selected using dark field optical microscopy, as tape residues can be easily spotted and avoided.

Resistivity values for van der Pauw devices are measured by applying a constant source-drain bias to two neighbouring corner contacts of the device and measuring the voltage drop between the opposite contacts during sweeping of the gate voltage. This permits two pairs (through exchanging the source-drain and measurement terminals) of independent measurements of the resistance of the device $R^{1,2}_{vertical}$ and $R^{1,2}_{horizontal}$ across the orthogonal 'vertical' and 'horizontal' directions (Supplementary Figure 7), and the pairs are then averaged to produce $R_{vertical}$ and $R_{horizontal}$. The sheet resistance $R_S$ is calculated by applying the reciprocal van der Pauw formula $\exp(-\pi R_{vertical}/R_S) + \exp(-\pi R_{horizontal}/R_S) = 1$.